\documentclass[a4paper,twoside,hidelinks]{article}

\usepackage[ngerman,main=english]{babel}

\usepackage{rotating}
\usepackage{tabularx}
\usepackage{enumitem} 
\usepackage{graphicx}
\usepackage{hyperref}
\usepackage{multirow}
\usepackage[table]{xcolor}    

\usepackage[textsize=footnotesize]{todonotes}
\usepackage{flushend} 

\usepackage{epsfig}
\usepackage{pslatex}
\usepackage{apalike}
\usepackage{algorithm2e}
\usepackage[bottom]{footmisc}

\usepackage[capitalise]{cleveref}

\usepackage{SCITEPRESS}     
\hypersetup{
    colorlinks,
    linkcolor={blue!70!black},
    citecolor={green!60!black},
    urlcolor={blue!70!black}
}

\newcommand{\msk}{msk}
\newcommand{\sk}{sk}

\begin{document}
\title{Protecting Privacy in Federated Time Series Analysis: \\ A Pragmatic Technology Review for Application Developers%
	\thanks{Authors are listed in alphabetical order, cf. also \url{https://www.ams.org/profession/leaders/CultureStatement04.pdf}.}}

	\author{
		\authorname{
			Daniel Bachlechner\sup{1}\orcidAuthor{0000-0001-7726-9065},
			Ruben Hetfleisch\sup{1}\orcidAuthor{0000-0002-9428-3835},
			Stephan Krenn\sup{2}\orcidAuthor{0000-0003-2835-9093},\\
			Thomas Lorünser\sup{2,3}\orcidAuthor{0000-0002-1829-4882},
			Michael Rader\sup{1}\orcidAuthor{0009-0000-1819-5246}
		}
		\affiliation{\sup{1}Fraunhofer Austria Research GmbH, Vienna / Wattens, Austria}
		\affiliation{\sup{2}AIT Austrian Institute of Technology GmbH, Vienna, Austria}
		\affiliation{\sup{3}Digital Factory Vorarlberg GmbH, Dornbirn, Austria}
		\email{\{daniel.bachlechner, ruben.hetfleisch, michael.rader\}@fraunhofer.at, \{stephan.krenn, thomas.loruenser\}@ait.ac.at}
	}


\abstract{
	The federated analysis of sensitive time series has huge potential in various domains, such as healthcare or manufacturing.
	Yet, to fully unlock this potential, requirements imposed by various stakeholders must be fulfilled, regarding, e.g., efficiency or trust assumptions.
	While many of these requirements can be addressed by deploying advanced secure computation paradigms such as fully homomorphic encryption, certain aspects require an integration with additional privacy-preserving technologies.
	\newline
	In this work, we perform a qualitative requirements elicitation based on selected real-world use cases.
	We match the derived requirements categories against the features and guarantees provided by available technologies.
	For each technology, we additionally perform a maturity assessment, including the state of standardization and availability on the market.
	Furthermore, we provide a decision tree supporting application developers in identifying the most promising technologies available matching their needs.
	Finally, existing gaps are identified, highlighting research potential to advance the field.
	}

\onecolumn \maketitle \normalsize \setcounter{footnote}{0} \vfill

\section{\uppercase{Introduction}}
The proliferation of data-driven technologies has led to an exponential increase in the collection and analysis of time series data across multiple domains.\footnote{\url{https://www.verifiedmarketresearch.com/product/time-series-analysis-software-market/}} Time series data, characterised by its sequential and temporal nature, provides critical insights that drive decision-making processes in sectors such as healthcare, manufacturing and smart infrastructure. However, the sensitivity of this data often includes personal, confidential or proprietary information, requiring strict privacy measures.

In healthcare, time series data from patient monitoring systems and electronic health records can reveal important trends and treatment outcomes, contributing significantly to medical research and patient care.
However, patient privacy must be protected to comply with regulatory standards such as HIPAA\footnote{\url{https://www.govinfo.gov/content/pkg/PLAW-104publ191/html/PLAW-104publ191.htm}} and GDPR\footnote{\url{https://eur-lex.europa.eu/legal-content/EN/TXT/?uri=CELEX:32016R0679}}, while still allowing for comprehensive data analysis. 
This delicate balance between data utilization and privacy is a significant challenge.

In the manufacturing sector, often referred to as Industry 4.0, time-series data collected from sensors and machines is essential for optimising production processes, predictive maintenance and improving operational efficiency. If this data is compromised, it can lead to significant competitive and operational risks. It is therefore critical to implement privacy technologies that can protect proprietary information while facilitating the analysis required for innovation and improvement.

Smart buildings, a specialised area of smart infrastructure that integrates IoT devices and sensors to monitor and control various systems such as lighting, heating and security, generate large amounts of time series data. This data can improve energy efficiency, enhance occupant comfort and ensure safety. However, it also contains sensitive information about the behaviour and preferences of building occupants, requiring robust data protection measures to prevent unauthorised access and misuse.

\medskip

What all these domains have in common is that valuable insights can be gained by combining data coming from different -- potentially mutually distrusting -- entities.
Thus, privacy-preserving technologies (PPTs) providing formal security guarantees are required to unlock the potential of the data without compromising confidentiality.
This is also in line with the paradigm of data sovereignty, as also pushed forward, e.g., by the European Data Spaces\footnote{\url{https://digital-strategy.ec.europa.eu/en/policies/data-spaces}}, guaranteeing users and organizations full control over their own sensitive data.

Besides social acceptance, the consideration of PPTs is also mandated by legal frameworks.
For instance, Article 32 GDPR requires to consider the available state of the art when selecting measures to ensure a level of security appropriate to the identified risks.
In this context, as clarified, e.g., by the European Union Agency for Cybersecurity (ENISA), a technology is considered as ``state of the art'' when it reaches market maturity, which, among others, depends on its availability and level of standardization.\footnote{\url{https://www.enisa.europa.eu/news/enisa-news/what-is-state-of-the-art-in-it-security}}

However, the landscape of PPTs is complex and rapidly evolving, making it challenging for both practitioners and researchers to navigate. Identifying technologies that have the necessary properties and are mature enough to be implemented in real-world scenarios remains a daunting task. This difficulty is compounded by the diversity of PPTs, ranging from secure multi-party computation to federated learning, each with its own strengths and limitations. As a result, there is an urgent need for comprehensive evaluations and guidelines to assist in the selection and use of the most appropriate technologies for specific time series analysis applications.

\subsection{Our Contribution}
This research explores the practical applicability of PPTs in real-world use cases, identifies research gaps, proposes a high-level research agenda and provides practical guidance for both practitioners and researchers aiming to implement PPTs in sensitive time series analysis.

More precisely, we provide developers of collaborative time-series analysis applications with a comprehensive overview of the available computing paradigms and complementary additional PPTs.
For each technology, we provide an analysis of its properties with respect to the most relevant requirement categories derived from real-world use cases, and support the developer by assessing the maturity level of each candidate.

Furthermore, we provide practical guidance in the identification of suitable candidate technologies for given scenarios.

Finally, we identify future research potential and propose a concise research agenda, to further increase the real-world applicability of available technologies.

\subsection{Related Work}
For an introduction to time-series analysis in general, we refer to the academic literature, e.g., \cite{brodav16}.

There are systematic overviews of PPTs and guidance for the selection of such technologies both in general~\cite{balase18} and for different application domains, e.g., for health data~\cite{jofohe22}, smart cities~\cite{PANDYA2023102987}, finance~\cite{DBLP:conf/aft/BaumCDF23}, or federated learning in telemedicine~\cite{hiwale2023} and production~\cite{sihera23}.
A comprehensive overview in the context of official statistics has recently been published by the United Nations Statistics Division~\cite{un-pets}.
Support in the maturity assessment was offered by ENISA.\footnote{\url{https://www.enisa.europa.eu/publications/pets}}\footnote{\url{https://www.enisa.europa.eu/publications/pets-maturity-tool}}

However, to the best of our knowledge, no comprehensive overview supporting developers in assessing the most promising technologies for specific application domains of (federated) time series exists.

\subsection{Paper Outline}
The paper is structured as follows: The results of a requirements analysis are presented in \Cref{sec:requirements}.
Advanced secure computation paradigms as well as complementary PPTs are introduced in \Cref{sec:technologies}, discussing their properties and maturity.
\Cref{sec:discussion} then discusses the pros and cons of all approaches and provides support for technology selection.
\Cref{sec:conclusion} concludes the paper with a research agenda and practical guidelines for practitioner and researchers.

\section{\uppercase{Requirement Categories}}\label{sec:requirements}
In the following we introduce a set of exemplary application scenarios in the context of federated time series analysis.
Supported by discussions with stakeholders from the domains, we performed a qualitative requirements elicitation for each of these uses.

The use cases were chosen from complementary domains to highlight the diverse needs and expectations associated with federated time series.
Although numerous specific use cases exist within each domain, the selected examples represent a wide spectrum of requirements, encompassing a broad range of applications in federated time series analysis.

Based on this, \Cref{sec:req_overview} then presents a clustering of the found requirements, showing clear similarities and differences in the needs of our use cases.

\subsection{Motivating Use Cases}\label{sec:use_cases}
\subsubsection{Domain 1: Medical Research}
This use case considers the scenario of multiple hospitals collaborating on joint medical studies. This is particularly relevant in the context of rare diseases, where a sufficiently large number of patients can only be monitored and analyzed by involving geographically distributed medical facilities.

A concrete example is related to post-operative pain management and the required amount of Fentanyl for treatment, as studied by Chiang et al.~\cite{smokers-fentanyl}, based on time series containing time stamps and dose.
Assuming that not all patients are treated in the same hospital, calculating the significance of different drug dosages can be challenging. Given that medical data is classified as special category data according to Article 10 of the GDPR, any computation or pooling of such information is constrained by stringent regulatory and ethical concerns.

\paragraph{Requirements.}
To facilitate such studies, it would be advantageous if data controlled by a hospital never leaves its facility, not even in encrypted form, as this still constitutes personally identifiable information.

Given that different studies typically involve various types of data and computations, a highly versatile setup is required. However, efficiency is only of secondary importance, and no real-time requirements exist. A reasonable IT infrastructure in terms of computational and bandwidth capacity is assumed.

Since hospitals are highly regulated, it can be assumed they will not maliciously deviate from protocol specifications. Nonetheless, because sensitive decisions (e.g., regarding treatment strategies) may depend on the outcomes, auditability would still be beneficial to the greatest extent possible.

\subsubsection{Domain 2: Industry 4.0}
Among others, Industry 4.0 aims at creating smart factories where machines, operators, and manufacturers are interconnected through the Internet of Things (IoT).
Machines generate vast amounts of data which they transmit to the Original Equipment Manufacturer (OEM).
The OEM uses this data to train sophisticated models aimed, e.g., at predictive maintenance and remote diagnostics.
This can significantly enhance operational efficiency, reduce downtime, and lead to cost savings, marking a substantial advantage over traditional production methods.

However, this continuous stream of data also brings substantial confidentiality concerns, as it may contain information about production processes or proprietary technologies.
This makes the OEM a central trusted party, from which sensitive information may leak, e.g., to competitors, leading to direct economic harm.

\paragraph{Requirements.}
What is thus needed are technologies that allow for the federated training of advanced models, without a central entity that learns any information about the underlying data.
Furthermore, in the optimal case, the different contributors to such a model do not need to know about each other, i.e., no direct communication among them is necessary.

Regarding efficiency requirements, no real-time needs exist for the training phase;
however, to avoid damage to the production machines, real-time analysis of the data is required.

The results of this data analysis step may have immediate consequences on business continuity:
in case of false negatives, unnecessary downtime may occur, increasing costs and reducing availability.
Thus, especially if inference is done remotely, any decision leading to a machine shutdown needs to be auditable by third parties to avoid fraud.

Additionally, competitors may have incentives to introduce malicious data into their data sets (also known as data poisoning), aiming for increasing downtime of their competitors.
As traditional countermeasures like data validation become harder in distributed settings without central authority, it would be desirable to receive cryptographic proofs that all data included in the training phase was indeed generated by a genuine sensor of the OEM.

\subsubsection{Domain 3: Smart Buildings}
Modern buildings are generating huge amounts of data, which can be used by property management, operators or manufacturers of certain devices (e.g., heating, cooling), or owners of the building.
However, while data access is clearly managed for certain areas, it becomes more complicated when residents require information about the data of other apartments:
for instance, it may be possible to optimize one's heating settings depending on the presence of neighbors, e.g., during winter holiday seasons.

Another illustrative example considers the following platform:
residents provide their energy consumption, and receive back feedback comparing their use (e.g., in form of a quantile) to that of their peers in the same building with comparable external conditions (e.g., whether, building insulation).
On the one hand such gamification could increase residents' willingness to reduce their energy consumption;
on the other hand this information could also be used to detect outliers, which might indicate needs for repair.

\paragraph{Requirements.}
Residents may be reluctant to disclosing detailed time series to a central authority, and especially to their peers, but they should rather remain in full control over their data.
On the other hand, however, usability is of utmost importance in such a setting, such that residents should not be required, e.g., to host compute nodes for federated computations themselves.
Thus, the computation should be offered as-a-service while guaranteeing privacy.

Assuming that feedback is provided, e.g., on a weekly or monthly basis, no real-time requirements exist.
However, given the computation is to be carried out repeatedly, pre-computations may be acceptable to increase efficiency.
Also, given the gamification approach, no immediate auditability or public verifiability requirements exist.

To minimize costs, no powerful IT infrastructure should be required to carry out the computations.

\subsection{Requirement Overview}\label{sec:req_overview}
Based on the scenarios described above we can now derive a set of generic requirements categories occurring describing the unique needs of each application case.
Following our ambition of providing guidelines also to non-experts in the fields in the specification of their applications and subsequently identifying the potential cryptographic approaches, these categories are described on a qualitative level to ease an initial assessment.

\begin{table*}[t!]
	\centering \footnotesize
	\setlength{\arrayrulewidth}{0.3mm}
	\setlength{\tabcolsep}{2pt}
	\renewcommand{\arraystretch}{1.33}
	\rowcolors{2}{white}{gray!25}
	\begin{tabularx}{\textwidth}{|p{2cm}|p{4.36cm}|p{4.36cm}|p{4.36cm}|}
		\cline{2-4}
		\rowcolor{white!50}
		\multicolumn{1}{c|}{}     & \multicolumn{1}{c|}{\bf UC1}                                                                                 & \multicolumn{1}{c|}{\bf UC2}                                                     & \multicolumn{1}{c|}{\bf UC3}             \\
		\rowcolor{white!50}
		\multicolumn{1}{c|}{}     & \multicolumn{1}{c|}{\bf Medical Research}                                                                    & \multicolumn{1}{c|}{\bf Industry 4.0}                                          & \multicolumn{1}{c|}{\bf Smart Buildings} \\
		\hline
		\bf Trust model           & Data should not leave own facility                                                                           & \multicolumn{2}{p{8.9cm}|}{Central data processor for protected data acceptable}                                            \\
		\hline
		\bf Scalability           & Specific computations per study                                                                              & \multicolumn{2}{p{8.9cm}|}{Medium to high frequency of same computation}                                                    \\
		\cline{2-4}
		                          & small data sizes                                                                                             & large data sizes                                                                 & medium data size                         \\
		\hline
		\bf Efficiency            & No specific requirements                                                                                     & Real-time needs for analysis                                                     & No specific requirements                 \\
		\hline
		\bf Flexibility           & Computations might not be known when generating data                                                         & \multicolumn{2}{p{8.9cm}|}{Computations clear upon data creation}                                                           \\
		\hline
		\bf Data integrity        & \multicolumn{2}{p{8.9cm}|}{Only authentic data should be included in computations; source privacy desirable} & No specific needs                                                                                                           \\
		\hline
		\bf Verifiability         & \multicolumn{2}{p{8.9cm}|}{Verifiability of computation results desirable}                                   & No auditing needs                                                                                                           \\
		\hline
		\bf Ease of integration   & \multicolumn{2}{p{8.9cm}|}{No inherent hardware limitations, skilled personnel may be available}             & Strong limitations regarding deployment, integration, costs                                                                 \\
		\hline
		\bf Regulatory contraints & Strict                                                                                                       & medium to strict (e.g. for critical infrastructure)                              & low to medium                            \\
		\hline
	\end{tabularx}
	\caption{Comparison of challenges for our selected use cases.}
	\label{tab:challenges_map}
\end{table*}

\begin{description}
	\item[Trust model.]
		This category describes the acceptable assumptions that may be made in an instantiation of the use case, thereby also describing data sovereignty needs.
		For instance, may data be processed (in an encrypted form) by a cloud provider, or does the computation need to happen in-house?
		Is there only a need to protect against accidental data leaks, or do active attacks by malicious parties need to be taken into consideration?
	\item[Scalability.]
		This category subsumes in particular the frequency of computations to be performed, the data sizes to be processed in each computation, or the number of users providing data.
	\item[Efficiency.]
		Actual efficiency requirements are specified in this category.
		This includes bandwidth and computational limitations, but also maximum execution times for each computation including real-time requirements.
	\item[Flexibility.]
		This category focuses on the computations to be performed, i.e., whether a single, static computation is computed periodically on different input data, or whether highly dynamic computations are to be supported.
		It also includes the function to be evaluated (e.g., Boolean or arithmetic circuit, multiplicative depth of circuits, etc.).
	\item[Data integrity.]
		This category is concerned with the quality of the processed data.
		In particular, it specifies whether input data needs to be checked for integrity and authenticity, which depends on the decisions to be taken based on the results, and on whether data providers are mutually trusting each other regarding the quality of the their data.
		Also, this category covers the aspect of completeness, meaning that special attention needs to be paid that no data is maliciously ignored during the computation.
	\item[Verifiability.]
		Closely related to the previous, this category covers all aspects related to quality of the computation result.
		That is, is it necessary that the receiver of a computation result obtains formal proofs that the computation was done correctly, or is the data processor trustworthy in the sense that is follows the protocol specification?
		Also, specific use cases may require public verifiability (or auditability), where also a third party is able to check the validity of the result.
	\item[Ease of integration.]
		This category covers all aspects related to available hardware and software systems, acceptable costs, or the availability of skilled personnel for design and maintenance of a system.
	\item[Regulatory constraints.]
		Finally, this class captures all known legal requirements, e.g., related to data protection or critical infrastructures that need to be fulfilled by any deployed system.
\end{description}

A qualitative mapping of these requirements to our use cases can also be found in \Cref{tab:challenges_map}.

\section{\uppercase{Technologies}}\label{sec:technologies}

In \Cref{sec:paradigms} we will now briefly introduce the different approaches found in the literature for federated computing in general and for time series in particular.
In \Cref{sec:extensions} we discuss additional PPTs that in combination with these paradigms can be used to overcome existing limitations.

\subsection{Secure Computing Paradigms}\label{sec:paradigms}
For each technology presented in the following, we provide a short description of their properties, as well as indications regarding its maturity and practical applicability.

\subsubsection{Secure Multi-Party Computation}
Secure multi-party computation (MPC) was first introduced by Yao~\cite{DBLP:conf/focs/Yao82b,DBLP:conf/focs/Yao86}, and subsequently generalized by Goldwasser et al.~\cite{DBLP:conf/stoc/GoldreichMW87} and further researched and developed in a large body of work, cf., e.g., \cite{DBLP:journals/ftsec/EvansKR18}.
In a nutshell, it allows parties to collaboratively perform any joint computation on their respective private inputs, without revealing any other information to the other parties than what is revealed by the predefined partial outputs to the individual parties.

The predominant approach used in efficient modern MPC protocols is the ``share of shares'' paradigm.
Initially, each party (or ``node'') holding a secret value runs a secret sharing scheme~\cite{DBLP:journals/cacm/Shamir79,krelor23} to generate shares of its secret, in a way that guarantees that only sufficiently large subsets of the shares can jointly gain any information about the secret.
The parties then distribute their shares to each other, such that each party ultimately holds one share of every other party's secret.
Based on those shares, the parties can perform computations on the joint data while maintaining privacy.
Finally, the nodes collaboratively reconstruct the final computation result by ensuring that only the shares of the final result are revealed to the receiver, but no shares of any intermediate results or other private data.

\paragraph{Properties.}
The privacy and integrity guarantees of MPC are based on a non-collusion assumption.
That is, these properties are maintained as long as no more than a predefined fraction of nodes collude, while no guarantees can be given if too many nodes collaborate maliciously.
In particular, distinctions are made depending on whether an honest majority is assumed or not.
Furthermore, different security notions exist, depending on whether malicious nodes still follow the protocol specification (``passive security'') or whether they may arbitrarily deviate (``active security'').
As a rule of thumb, the efficiency of MPC protocols decreases with the strength of the attacker model, resulting in a sometimes delicate balance between achieving the desired levels of security and efficiency.

However, despite its versatility, MPC also comes along with several limitations.
Specifically, while any computation on sensitive inputs can be performed using MPC, the structure of the computation to be performed strongly influences the efficiency of the resulting protocol.
For instance, for protocols based on Shamir's secret sharing scheme~\cite{DBLP:journals/cacm/Shamir79}, additions are basically for free, while multiplications require complex, interactive operations among the nodes, potentially causing significant delays depending on network latency and bandwidth.
Therefore, circuits with low multiplicative depth are preferable in this approach.
Furthermore, the bandwidth requirements often heavily depend on the size of the secret inputs held by the different parties, such that applications with small(er) input sizes are often more suitable for MPC than such with big input sizes.

Finally, depending on the concrete setup, MPC can be a useful technical control to (partially) overcome the challenges imposed by data protection regulations~\cite{10.1007/978-3-031-09901-4_2}.

\paragraph{Maturity level.}
MPC is a highly versatile tool that has been used to demonstrate a variety of practical application domains, ranging from secure auctions \cite{DBLP:conf/fc/BogetoftCDGJKNNNPST09} over industry 4.0 \cite{DBLP:conf/icissp/LorunserWK22} to finance \cite{DBLP:conf/fc/GamaCPSA22}.

Usable frameworks supporting the development of MPC protocols exist, e.g., MP-SPDZ~\cite{DBLP:conf/ccs/Keller20}.
Furthermore, several standards developing organizations (SDOs) have already published standards on MPC, including ISO/IEC (within its ISO/IEC 4922 series\footnote{\url{https://www.iso.org/standard/80508.html}}) and IEEE SA\footnote{\url{https://standards.ieee.org/ieee/2842/7675/}}.

Guidelines and open challenges for the integration of MPCs into data infrastructures such as European Data Spaces have recently been investigated by Siska et al.~\cite{DBLP:conf/closer/SiskaLKF24}.

\subsubsection{Fully Homomorphic Encryption}
While in traditional encryption schemes basically the only thing that can be done with a ciphertext is to decrypt it, fully homomorphic encryption (FHE) allows one to perform arbitrary computations on ciphertexts in the encrypted domain.
Thus, FHE ensures that the operations on the ciphertext correspond to well-defined operations on the underlying plaintext data.
This enables data owners, e.g., to encrypt their data before sending them to a powerful cloud service.
The cloud provider may then perform computations (e.g., data analysis on sensitive data) on the ciphertexts without ever having access to the plaintext data.
Eventually, the cloud provider will return the encrypted result to the holder of the secret key who only needs to decrypt the final result.

The concept of FHE was first envisioned by Rivest et al.~\cite{Rivest1978} already in 1978, but only first instantiated in 2009 by the seminal work of Gentry~\cite{DBLP:conf/stoc/Gentry09}.
It has been a highly active area of research ever since, resulting in significant speedups in terms of ciphertext sizes and computational costs.

\paragraph{Properties.}
Evaluating functions using FHE is several orders of magnitude slower compared to a plaintext evaluation~\cite{DBLP:journals/corr/abs-2202-02960}.
In particular, after a certain number of operations (e.g., multiplications), highly expensive so-called ``bootstrapping'' steps need to be performed to guarantee valid decryptions.
This is partially overcome by somewhat homomorphic encryption schemes~\cite{DBLP:conf/crypto/Brakerski12}, which support only a small number (e.g., one) of multiplications but arbitrarily many additions.
They thereby avoid the need for bootstrapping steps while limiting the expressiveness of the schemes.
Thus, such schemes can significantly increase efficiency for specific application cases.

If used to analyze data coming from different data sources, special attention needs to be paid to the management of the master secret key $\msk$, as this could not only be used to decrypt the computation results, but also the encrypted inputs.
Thus, in scenarios where the intended receiver of the computation result (owning $\msk$) must not get access to the individual encrypted inputs, it needs to be ensured that the computing server, e.g., does not leak encrypted input data to the receiver.
This can be achieved on an organization level, and by enforcing strict access policies.
Alternatively, multikey-FHE schemes, first introduced by~\cite{DBLP:conf/stoc/Lopez-AltTV12}, are capable of computing on ciphertexts encrypted under multiple unrelated keys can be deployed, yet causing higher computational costs and additional complexity as all individual secret keys are required for decryption.

Finally, verifiability of the performed computation can be achieved by deploying specific schemes, e.g.,~\cite{DBLP:journals/corr/abs-2301-07041}.

\paragraph{Maturity level.}
Similar to MPC, 
FHE is a very flexible technology that can be used to perform arbitrary computations on sensitive data, which has been proved in a variety of scenarios including federated learning~\cite{DBLP:journals/iacr/ZhaoZW24} or healthcare~\cite{munbha22}.

Over the last years, FHE has made significant advancements regarding practicability, and a range of open-source frameworks are available to researchers and developers\footnote{\url{https://www.zama.ai/}}$^,$\footnote{\url{https://github.com/homenc/HElib}}$^,$\footnote{\url{https://www.openfhe.org/}}.
Finally, fully homomorphic encryption is subject to ongoing standardization efforts, e.g., by ISO/IEC (within the ISO/IEC 28033 series\footnote{\url{https://www.iso.org/standard/87638.html}}), ITU-T\footnote{\url{https://www.itu.int/ITU-T/workprog/wp_item.aspx?isn=17999}}, and community standardization efforts\footnote{\url{https://homomorphicencryption.org/}}.
Finally, in order to overcome efficiency limitations, substantial efforts are currently taken to provide hardware acceleration for FHE in the near future.\footnote{\url{https://www.darpa.mil//news-events/2021-03-08}}\footnote{\url{https://dualitytech.com/partners/intel/}}

\subsubsection{Functional Encryption}
Similar to FHE, functional encryption (FE) \cite{DBLP:conf/tcc/BonehSW11} offers a more versatile approach than all-or-nothing access to encrypted data, by allowing the holder of the master secret key $\msk$ to generate partial decryption keys $\sk_f$.
Now, upon receiving a ciphertext that encrypts a message $m$, the holder of such a partial decryption key cannot recover the message itself, but only $f(m)$, where $f(\cdot)$ is a function defined by the holder of the master secret key.
If the key is controlled, e.g., by the data subject, fine grained access to computations on sensitive data can be provided.

Generalizations to schemes that can evaluate functions on multiple inputs -- i.e., $\sk_f$ allows for computing $f(m_1,\dots,m_\ell)$, where each $m_i$ is encrypted in an individual ciphertext -- were presented by Goldwasser et al.~\cite{DBLP:conf/eurocrypt/GoldwasserGG0KLSSZ14}.
Other generalizations include, e.g., schemes where the inputs are not necessarily coming from the same entity, allowing for computations across data from different data subjects~\cite{DBLP:conf/asiacrypt/ChotardSGPP18,DBLP:conf/asiacrypt/NguyenPP23}.

\paragraph{Properties.}
While requiring proper protection of the master secret key $\msk$ to ensure confidentiality of all data, functional encryption provides an interesting trust model, as the decryption -- and thus the computation -- is directly carried out by the receiving party, reducing the need for verifiability.

On the downside, FE is only meaningful in scenarios where the function to be computed is known in advance, as the schemes need to support the right class of functions.
In particular, most efficient FE schemes are focusing on the computation of inner products (including, e.g., weighted sums of data items).

\paragraph{Properties.}
Functional encryption has been proven useful in different contexts such as, e.g., biometric authentication~\cite{DBLP:journals/tetc/AdhikaryK24} or machine learning~\cite{DBLP:journals/iotj/PanzadeT024}.

Libraries for functional encryption are available in different languages through, e.g., the FENTEC project\footnote{\url{https://github.com/fentec-project}} of by Adhikary and Karmakar\footnote{\url{https://github.com/s-adhikary/IPFE}}.

To the best of our knowledge, there are no ongoing efforts for standardizing functional encryption.

\subsubsection{Trusted Execution Environments}
Trusted Execution Environments (TEEs) are secure areas within a processor designed to protect sensitive code and data from unauthorized access or modification by the operating system or other applications~\cite{Sabt.2015}. TEEs function by providing isolated execution spaces that operate alongside the main operating system, ensuring an additional layer of security for sensitive processes. In practice, TEEs work by offering a restricted processing environment where only authorized code can execute, and memory is encrypted to prevent external inspection or tampering. This is critical for handling operations such as cryptographic key management, biometric data processing, and secure financial transactions~\cite{Lind.2016}. When an application needs to perform a secure operation, it makes a call to the TEE, which processes the request in isolation, safeguarding the confidentiality and integrity of the data being handled. This makes TEEs vital for enforcing privacy and security in a broad range of digital applications.

\paragraph{Properties.}
This is also essential for compliance with stringent data protection regulations in industries like finance and healthcare. Additionally, TEEs help in mitigating a range of security threats, including malware and software vulnerabilities. Their ability to handle secure transactions and authenticate user interactions in a protected manner makes them invaluable for mobile banking, DRM (Digital Rights Management), and personal identification applications. Also the integrity and authenticity of code execution can be reliably verified ~\cite{Chen.2023}. Overall, TEEs strengthen the trustworthiness of devices and systems, boosting user confidence in digital platforms.
TEEs face several challenges that can impact their broader application and effectiveness. One of the primary hurdles is hardware dependency, as TEEs require specific processor features to function, which can limit their use to certain devices and platforms, thus restricting scalability~\cite{Geppert.2022}. Integrating TEEs into existing systems also poses significant challenges due to the need for specialized knowledge to handle secure communication and operation management between the TEE and the rest of the system. 
This process is further complicated, as any code loaded into secure enclaves needs to be signed using tools provided by the TEE manufacturer to ensure security, making quick adaptations of the code complicated. 
Moreover, the complexity of developing and maintaining secure code that runs within TEEs increases the risk of vulnerabilities if not properly managed.
Another main point to consider are strong trust assumptions that need to be put into the trustworthiness of the hardware manufacturer.

Additionally, cross-platform compatibility issues can arise, complicating the development of applications that can operate across different types of devices and operating systems with TEEs. 

Finally, the evolving landscape of cybersecurity threats continuously tests the resilience of TEEs, necessitating ongoing updates and patches to maintain a robust defense.
While this is true for any security-related implementation -- e.g., considering side-channel attacks --, this issue requires further attention due to the required trust in hardware components:
in case that an attack exploits hardware vulnerabilities, patches may be more complex and cost intensive than in the case of pure software applications. 

\paragraph{Maturity level.}
Trusted Execution Environments are provided by several companies, often integrated into their hardware or software products, including, e.g., Intel's Software Guard Extensions (SGX)\footnote{\url{https://www.intel.com/content/www/us/en/products/docs/accelerator-engines/software-guard-extensions.html}}, AMD Secure Encrypted Virtualization (SEV)\footnote{\url{https://www.amd.com/en/developer/sev.html}}, ARM TrustZone\footnote{\url{https://www.arm.com/technologies/trustzone-for-cortex-m}}, or IBM Secure Service Container\footnote{\url{https://www.ibm.com/support/pages/ibm-z-secure-service-container-users-guide}}, just to name a few.

The standardization of TEEs has been primarily driven by the GlobalPlatform organization\footnote{\url{https://globalplatform.org/}}.

\subsubsection{Federated Learning}
Federated Learning (FL) is a collaborative machine learning approach, first introduced by McMahan et al.~\cite{DBLP:conf/aistats/McMahanMRHA17}.
It enables model training without centralizing data, thus preserving the privacy and security~\cite{Konecny.18.10.2016} of training data. Therefore, it allows organizations to jointly develop models without sharing sensitive data, fostering international collaboration and building more generalizable models~\cite{Kiss.2022}. Instead of pooling data on a central server, each participant trains a model on their own device using their private data and then sends only the model updates (weights or gradients) to a central server. The central server aggregates these updates to improve a global model, which is then sent back to the participants for further training. This cycle repeats, enhancing the global model with every step.

To avoid, e.g., model inversion or model inference attacks, typically a trusted server is assumed for aggregation, but also secure aggregation protocols exist \cite{DBLP:conf/ccs/BonawitzIKMMPRS17}.

\paragraph{Properties.}
Federated learning offers significant advantages, chiefly in privacy preservation, as it allows data to remain on local devices, reducing the risk of data breaches \cite{Li.2022}. Furthermore, the data security of FL is usually supplemented with differential privacy to further protect potentially sensitive training data \cite{9069945}. This decentralized approach also enhances scalability by distributing computation across numerous devices, thus avoiding centralized processing bottlenecks. Additionally, it supports data diversity, reflecting real-world variations, and improves model robustness. By leveraging multiple data sources without direct access, it complies with data protection regulations, making it suitable for industries like healthcare and finance where data sensitivity is paramount. Overall, federated learning enables more secure and scalable machine learning deployments \cite{Zhang.2021}.

Despite its advantages, FL also faces several limitations.
In the case of non-iid data distributions (data points are not independently and identically distributed, but may, e.g., be correlated), the 'client-drift' problem may cause clients to converge to local optima, leading to slower overall convergence and poor model performance \cite{Moshawrab.2023}. Another limitation lies in the difficulty of determining hyperparameter values, hindering the global model's performance in each update; a proposed solution involves dynamically adjusting hyperparameters using a log function to address this issue.

Integration of FL into complex systems comes with a variety of challenges.
Varying computational capabilities of participant devices, the complexity of establishing reliable communication channels, or the need to adapt existing systems to handle decentralized data processing need to be addressed.
At the same time, the consistent and secure participation of joining and exiting nodes in the network must be addressed~\cite{Moshawrab.2023}.

\paragraph{Maturity level.}
Federated learning is a rapidly advancing technique, which has found practical applications in diverse domains such as healthcare \cite{DBLP:journals/comcom/MuazuMMIKS24}, finance \cite{DBLP:conf/cdc/ShiSX23}, and mobile edge networks \cite{DBLP:journals/comsur/LimLHJLYNM20}.

Usable frameworks and platforms, such as TensorFlow Federated\footnote{\url{https://www.tensorflow.org/federated}}, Flower\footnote{\url{https://flower.ai/}}, and PySyft\footnote{\url{https://github.com/OpenMined/PySyft}}~\cite{Ziller2021}, support the development and deployment of federated learning protocols. Finally, several SDOs are actively working on federated learning standards, with the IEEE publishing standards such as IEEE P3652.1\footnote{\url{https://standards.ieee.org/ieee/3652.1/7453/}}.

\subsection{Extensions}\label{sec:extensions}
In the following we give a concise overview of possible extensions that may be integrated to achieve all the requirements identified in \Cref{tab:challenges_map}.

\subsubsection{Advanced Signature Schemes}
As described in \Cref{sec:requirements}, the integrity and authenticity of the data processed in certain computations can be crucial, in particular when sensitive decisions are to be taken depending on the outcome.
For instance, in the medical domain, and especially in scenarios where data is generated directly by patients, it might be desirable that only authentic data indeed generated by a genuine device is taken into consideration, in order to reduce the risk of incorrect or manipulated data.
For instance, only data coming from a genuine glucometer might be used in a diabetes study.
However, in order to protect the privacy of the single patient, simply having the data signed by the device is often not a solution, as this would allow for uniquely assigning measurements to individual patients.

This can be overcome, e.g., by the use of group signatures~\cite{DBLP:conf/eurocrypt/ChaumH91}.
In such a scheme, a group manager (e.g., the device manufacturer) holds a master key, and equips each user (e.g., each glucometer) with a separate and independent signing key.
Using this signing key, the user can now sign any data of her choice (e.g., encrypted blood sugar measurements).
The receiving party (e.g., the hospital server processing the measurements) can now use the group manager's public key to verify the authenticity of the received data, without learning which user actually generated the data.
That is, the verifier learns that the data has been generated by \emph{some} eligible user (i.e., a genuine glucometer), but not by \emph{which} user (i.e., patient).
Yet, in the case of anomalies or abuse of privacy, a dedicated third party (e.g., an ethics board) can de-anomymize a user and reveal her identity.

Similar privacy guarantees for more ad-hoc settings can be achieved by the use of ring signatures~\cite{DBLP:conf/asiacrypt/RivestST01}, which are also heavily used to achieve privacy in crypto currencies.

\paragraph{Maturity level.}
All the aforementioned technologies are highly mature and have been included into relevant standards, or are currently subject to standardization, e.g., within ISO/IEC (as part of the ISO/IEC 20008 series\footnote{\url{https://www.iso.org/standard/57018.html}}), cf.~\cite{krpotr22}.

Practical applications of group signatures include, e.g., direct anonymous attestation (DAA) implemented in Trusted Platform Modules (TPMs) as specified by the Trusted Computing Group\footnote{\url{https://trustedcomputinggroup.org/}}.

\subsubsection{Zero-knowledge proofs}
Most of the computing paradigms mentioned in \Cref{sec:paradigms} do not offer verifiability of the computation result by default.
One possibility to achieve this feature is to deploy so-called zero-knowledge proofs of knowledge (ZKPs)~\cite{DBLP:conf/stoc/GoldwasserMR85}, including more compact versions such as zero-knowledge succinct non-interactive arguments of knowledge (zk-SNARKs)~\cite{DBLP:conf/eurocrypt/Groth16}.
These are two-party-protocols between a prover and a verifier, which allow the former to prove knowledge of a certain piece of information, without disclosing anything beyond what is already revealed by the claim itself.

For instance, in the context of MPC, the compute nodes could jointly compute a zero-knowledge proof that proves that the nodes possess input data that has $(i)$ been signed by the data providers, such that $(ii)$ this data has been used in the defined computation, and $(iii)$ the provided output indeed corresponds to this computation result.
The feasibility of this approach has been proven in the contexts, e.g., of smart manufacturing~\cite{DBLP:conf/icissp/LorunserWK22} or air traffic management~\cite{DBLP:conf/ccsw-ws/LorunserWK22}.

A special application of zero-knowledge proofs is that of authentic inference from neural networks.
That is, in some applications on entity (e.g., an OEM) holds a high-quality neural network which they wish to protect, while customers want to make sure that indeed this network was used to obtain an inference result.
Using zero-knowledge proofs it is possible to prove that a specific network was used, without disclosing the network itself.
A first important step towards practical zero-knowledge inference is given by zkCNN~\cite{DBLP:conf/ccs/LiuXZ21}.

\paragraph{Maturity level.}
ZKPs have reached a relatively mature stage, with several practical implementations and real-world applications, including, e.g., private transactions in blockchains%
such as in Zcash\footnote{\url{https://z.cash/}} and layer-2 scaling solutions like zk-rollups in Ethereum\footnote{\url{https://ethereum.org/en/developers/docs/scaling/zk-rollups/}}.

Companies like StarkWare, zkSync, and Aztec are leading in practical implementations of ZKPs.
Standardization of zero-knowledge proofs is advancing, driven, e.g., by community standardization efforts such as ZKProof\footnote{\url{https://zkproof.org/}}.

\subsubsection{Advanced Encryption Mechanisms}
Consider a system where certain computations are performed on a periodic basis, and end users can purchase subscriptions to these periodic outcomes.
This may lead to asynchronous scenarios, where eligible receivers are not involved or even online at the time the computation is carried out and additionally change over time, such that dynamic access control mechanisms on the computation result are required.

Traditional solutions deploying, e.g., a central policy-enforcement point deciding whether a user has access or not, suffer from the drawback that the result needs to be stored in plaintext and only eligible users are granted access.
However, depending on the application domain, this may be undesirable from a data protection point of view.
On the other hand, encrypting the result under a traditional public-key encryption scheme individually for each user does not scale either in case of many potential data receivers.

An alternative mechanism is based on advanced encryption mechanisms such as attribute-based encryption~\cite{DBLP:conf/eurocrypt/SahaiW05,DBLP:journals/iotj/RasoriMPD22}.
In such a scheme, users can obtain secret keys linked to some attributes (e.g., the validity period of their subscription).
Now, instead of directly outputting the computation result, the compute nodes in any of the above mechanisms could output a ciphertext encrypted under a central public key, labeled with an access policy (e.g., the creation date of the ciphertext), and make the ciphertext publicly available.
Now, while anybody may access the ciphertext, only users with matching secret keys (i.e., where the creation date of the ciphertext is within the subscription period) can decrypt the contained information.

If the receiving party is known upfront but not online at the computation time, or if the data receiver in an FHE scenario is not equal to the holder of $\msk$, alternative encryption paradigms such as proxy re-encryption \cite{DBLP:conf/eurocrypt/BlazeBS98,DBLP:conf/asiacrypt/ZhouLHZ23} to translate ciphertexts into encryptions for the dedicated receiver without revealing the plaintext exist.

In either case, access management is enforced on a fully cryptographic level, without plaintext results ever being available to unprivileged entities.

\paragraph{Maturity level.}
The maturity of ABE is steadily increasing, with growing interest and several practical implementations but not yet widespread adoption.
Yet, several companies like Virtru\footnote{\url{https://www.virtru.com/}} implement ABE in their security solutions.
Furthermore, several open-source implementations are available~\cite{DBLP:journals/corr/abs-2209-12742}.

Standardization efforts are ongoing, including, e.g., ETSI's TS 103~532\footnote{\url{https://portal.etsi.org/webapp/workprogram/Report_WorkItem.asp?WKI_ID=62733}}.

\section{\uppercase{Discussion}}\label{sec:discussion}

\begin{table*}[th!]
	\centering \footnotesize
	\setlength{\arrayrulewidth}{0.3mm}
	\setlength{\tabcolsep}{2pt}
	\renewcommand{\arraystretch}{1.33}
	\rowcolors{2}{white}{gray!25}

	\begin{tabularx}{\textwidth}{ |p{2.00cm}|p{2.548cm}|p{2.548cm}|p{2.548cm}|p{2.548cm}|p{2.548cm}| }
		\cline{2-6}
		\multicolumn{1}{c|}{}      & \multicolumn{1}{c|}{\bf MPC}                                                                                         & \multicolumn{1}{c|}{\bf FHE}                                                                                                                  & \multicolumn{1}{c|}{\bf FE}                                                                                         & \multicolumn{1}{c|}{\bf FL}                                                                        & \multicolumn{1}{c|}{\bf TEE}                                                                    \\
		\hline
		\bf Trust model            & Non-collusion assumptions; support of active and passive adversary models                                            & Computations performed by untrusted data processor; confidentiality depends on proper management of secret key                                & Computations performed by data receiver; requires proper protection of master secret key                            & Trust in data owners to use righteous data; requires aggregator to protect raw data                & Hardware-based trust anchor; TEE provider needs to be trusted                                   \\
		\hline
		\bf Scalability            & \multicolumn{3}{p{7.6cm}|}{Well suited for small to medium data sizes}                                               & \multicolumn{2}{p{5.096cm}|}{Well suited also for large data sets}                                                                                                                                                                                                                                                                                                                                                                                                         \\
		\hline
		\bf Efficiency             & Communicational costs increase with the complexity of the circuit to be evaluated                                    & Non-interactive. Computational costs increase with complexity of the circuit                                                                  & Non-interactive. Costs depend on computation to be computed and scenario (input from one or multiple parties, etc.) & Computational costs decreased through local lightweight models but increased communicational costs & Moderate overhead compared to plaintext computation                                             \\
		\hline
		\bf Flexibility            & \multicolumn{2}{p{5.cm}|}{Support of arbitrary functions; complexity increases with multiplicative depth of circuit} & Limited, as type/class of computations need to be known upfront                                                                               & Support of a broad range of FL methods; increased communicational costs with increased complexity of functions      & Support of arbitrary functions; complex administrative processes by TEE provider                                                                                                                     \\
		\hline
		\bf Data integrity         & \multicolumn{5}{c|}{None by default, cf. \Cref{sec:extensions}}                                                                                                                                                                                                                                                                                                                                                                                                                                                                                                                                   \\
		\hline
		\bf Verifiability          & Resilience against a certain number of malicious nodes can be achieved. Publicy verifiable protocols exist.          & Not by default, but specific verifiable FHE schemes exist                                                                                     & None by default, yet computation can be carried out be data receiver, reducing the need for verifiability           & None by default, but impact of data quality traceable                                              & Integrity based on the trust assumptions into the secure enclave                                \\
		\hline
		\bf Ease of integration    & If hosted on-premise, computing infrastructure and trained personnel required.                                       & Low integration costs if used as-a-service, initial setup may require trained personnel. Key management may require organizational processes. & Low to moderate. Key management may require organizational processes.                                               & Complex Integration due to setting up infrastructure between nodes                                 & Trained personnel needed in development phase; complex organizational processes to certify code \\
		\hline
		\bf Regulatory constraints & \multicolumn{5}{c|}{Requires analysis on a case-by-case basis}                                                                                                                                                                                                                                                                                                                                                                                                                                                                                                                                    \\
		\hline
	\end{tabularx}
	\caption{Comparison of guarantees offered by different computing paradigms}
	\label{tab:computing_paradigms}
\end{table*}

Based on the descriptions in \Cref{sec:paradigms}, it can be seen that the available computing paradigms address the requirements set out in \Cref{tab:challenges_map} in highly orthogonal ways, and that some of those requirements are already effectively addressed by existing technologies.
We provide an overview of the properties of the computing paradigms in \Cref{tab:computing_paradigms}.

\smallskip

For instance, for the underlying trust model, a broad range of options is available, ranging from  non-collusion assumptions in keyless and information-theoretically secure settings (MPC) over computational assumptions provided by encryption schemes with central key material (FHE, FE), up to hardware-based trust anchors (TEE).
Regarding public verifiability and integrity of the computation result, TEEs and MPC can already provide certain guarantees by ensuring that computations on sensitive data are executed securelyand correctly.
For scalability, especially FL and TEEs provide efficient solutions that maintain reasonable performance levels, with FL excelling at distributing the computational load to address scalability concerns.
In addition, FL and TEEs offer considerable flexibility and ease of integration, making them suitable for diverse applications across multiple domains.

\smallskip

However, some requirements can not yet be fully achieved by certain available solutions.
For instance, FHE and MPC, while providing strong privacy guarantees, face challenges in efficiently scaling with large datasets due to high computational and communication overheads.
These technologies also significant efficiency improvements to be practical for real-time applications.
Ensuring the verifiability of computations without compromising privacy remains a challenge and requires improvements to provide stronger and more practical verifiability mechanisms for most approaches, including FE or FL.
Although flexible, FL needs further development to handle highly heterogeneous data sources and complex time series data structures.
Regarding integration efforts, most technologies require a certain level of skilled personnel at least for the initial setup.
The infrastructure requirements are relatively low if deployed as-a-service, yet can increase significantly if the computations are to be carried out locally on premise, which however also increases the degree of data-sovereignty.

\smallskip

Finally, certain requirements are not immediately addressed by any of the technologies.
For instance, none of the computing paradigms inherently supports data integrity validation, but they all assume valid input data.
For all solutions, a legal analysis has always to be carried out on a case-by-case basis, including further dimensions such as the criticality of the data or additional safeguards implemented.

Finally, for many of the paradigm-specific or general limitations, additional compatible PPTs exist, as also discussed in \Cref{sec:extensions}.

\medskip

\begin{figure}[th!]
	\centering
	\includegraphics[width=\columnwidth]{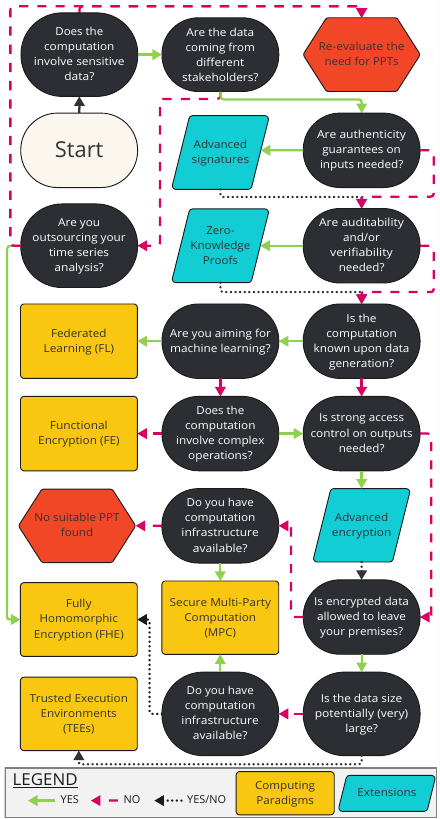}
	\caption{Decision tree supporting the selection of appropriate PPTs for federated time series analysis}
	\label{fig:decisiontree}
\end{figure}

Finally, in \Cref{fig:decisiontree} we provide a decision diagram supporting developers in efficiently identifying suitable candidates for PPTs in the context of federated time series analysis.

\section{\uppercase{Conclusion}}\label{sec:conclusion}

This research highlights the critical need for robust PPTs in the federated analysis of sensitive time-series data across different domains. Each considered domain -- healthcare, manufacturing and smart infrastructure -- presents unique challenges and requirements. Our comprehensive evaluation of advanced PPTs and selected extensions reveals their respective strengths and limitations in meeting these requirements.

We recognize several research gaps, including the need for more efficient algorithms for FHE and MPC to reduce computational and communication overheads. Improving the verifiability of computations in privacy-preserving settings is crucial to ensure that results can be independently verified without compromising privacy. Improving the interoperability of different PPTs is essential to facilitate seamless integration and transition between methods as required by specific use cases. In addition, the development of optimized solutions tailored to the specific needs of each domain, especially when dealing with time series data, is of paramount importance.

To address these gaps, future research should focus on algorithmic innovation, efficiency benchmarking, cross-technology integration, domain-specific solutions and regulatory alignment. Investment in the development of new algorithms and protocols that improve the efficiency and scalability of FHE and MPC is essential. The establishment of comprehensive benchmarking frameworks to systematically evaluate the performance of different PPTs in real-world scenarios will provide valuable insights. Focusing on the integration of multiple privacy-preserving technologies to exploit their combined strengths and mitigate individual weaknesses will enhance their practical applicability. Conducting in-depth studies and developing tailored solutions for the specific requirements of critical domains, with a particular focus on the challenges of time series data, will address domain-specific needs. Finally, engagement with regulators to ensure that emerging technologies are aligned with evolving data protection regulations and provide practical guidance on compliance will be essential for wider adoption.

By addressing these research gaps and pursuing the proposed research agenda, significant progress can be made in the field of privacy-preserving time series analysis. These advances will ultimately enable secure and efficient federated analysis of sensitive data across critical domains.

	\section*{\uppercase{Acknowledgement}}
	This work was funded by the Austrian Research Promotion Agency FFG within the PRESENT project (grant no. 899544). Views and opinions expressed are however those of the authors only and do not necessarily reflect those of the funding agency. Neither the FFG nor the granting authority can be held responsible for them.

	\bibliographystyle{apalike}
\bibliography{bibfile}

\end{document}